\documentclass[%
 reprint,
 amsmath,amssymb,
 aps,showpacs,prl,letterpaper,braket, superscriptaddress
]{revtex4-1}
\usepackage{braket}
\usepackage{graphicx}
\usepackage{dcolumn}
\usepackage{bm}
\usepackage{epstopdf}
\usepackage{comment}
\usepackage{float}

\begin{document}


\title{How to dress radio-frequency photons with tunable momentum}

\author{Boris Shteynas}
\thanks{These two authors contributed equally.}
\author{Jeongwon Lee}
\thanks{These two authors contributed equally.}
\author{Furkan~\c{C}a\u{g}r{\i}~Top}
\author{Jun-Ru Li}
\author{Alan O. Jamison}
\affiliation{Research Laboratory of Electronics, MIT-Harvard Center for Ultracold Atoms, Department of Physics,
Massachusetts Institute of Technology, Cambridge, Massachusetts 02139, USA}
\author{Gediminas Juzeli\ifmmode \bar{u}\else \={u}\fi{}nas}
\affiliation{Institute of Theoretical Physics and Astronomy, Vilnius University, Saul\.etekio 3, Vilnius 10257, Lithuania}
\author{Wolfgang Ketterle}
\affiliation{Research Laboratory of Electronics, MIT-Harvard Center for Ultracold Atoms, Department of Physics,
Massachusetts Institute of Technology, Cambridge, Massachusetts 02139, USA}

\date{\today}

\begin{abstract}
We demonstrate how the combination of oscillating magnetic forces and radio-frequency (RF) pulses endows RF photons with tunable momentum.  We observe velocity-selective spinflip transitions and the associated Doppler shift. This realizes the key component of purely magnetic spin-orbit coupling schemes for ultracold atoms, which does not involve optical transitions and therefore avoids the problem of heating due to spontaneous emission.
\end{abstract}


\maketitle


The field of cooling and trapping atoms depends on mechanical forces exerted by light through photon recoil \cite{PA}. Since photons can be scattered only by admixing electronically excited states, the mechanical forces due to light always involve some amount of dissipation by spontaneous emission. This is desirable in laser cooling but causes heating and atom loss in other situations where, it is often suppressed by using far off resonant light (e.g., in optical lattices).

The latter applies to recent efforts to create spin-orbit coupling \cite{Galitski2013, Hui, Zhang2018, SpielmanSOC, HuiExp, Martin} and synthetic gauge fields for ultracold atoms \cite{DalibardGauge, Goldman2014, Jaksch, SpielmanSynthetic, Hiro, Monica}, motivated by the goal of quantum simulations of new forms of matter.  However, when spin orbit-coupling is realized in alkali atoms with a two-photon process \cite{SpielmanSOC, Martin, HuiExp}, spontaneous emission cannot be suppressed by detuning since the strength of spin-orbit coupling and spontaneous scattering scale with detuning in the same way \cite{Hui}.

This limitation has motivated the development of alternative schemes of spin-orbit coupling.  Several groups have demonstrated \cite{Ketterle, Li2017, Engels} or suggested \cite{Sun2016} spin-orbit coupling with orbital states in optical lattices. This allowed spin-orbit coupling by a two-photon Raman process where spontaneous emission could be completely suppressed by far detuning. If the two coupled spin states have different magnetic moments, spin-orbit coupling can be induced without optical photons by time-dependent magnetic fields, and several schemes have been proposed \cite{Gediminas, Ueda, Luo2016, foot}.

In this work, we demonstrate the key element of these magnetic spin-orbit coupling schemes.  We drive RF transitions between two different hyperfine states in the presence of an alternating magnetic field gradient.  The time-averaged evolution is  an RF transition where recoil momentum is transferred.  The sign and magnitude of the momentum kick is adjustable via the magnetic fields, and we observe a recoil momentum which is $6 \times 10^6$ higher than the (usually negligible) momentum of an RF photon around 8 MHz frequency.

Our scheme shows the power of Floquet engineering:  we combine an RF transition, which has negligible momentum transfer, with a sinusoidally oscillating magnetic field gradient, which has no time-averaged momentum transfer, and the result is an RF photon with recoil, depending on how RF pulses are synchronized with the time-dependent magnetic field gradient.  This scheme is conceptually very transparent and illustrates important elements of Floquet physics, as well as the role of mechanical and canonical momenta in implementing synthetic gauge fields.

\begin{figure}[h]
\includegraphics[width = 8.6cm]{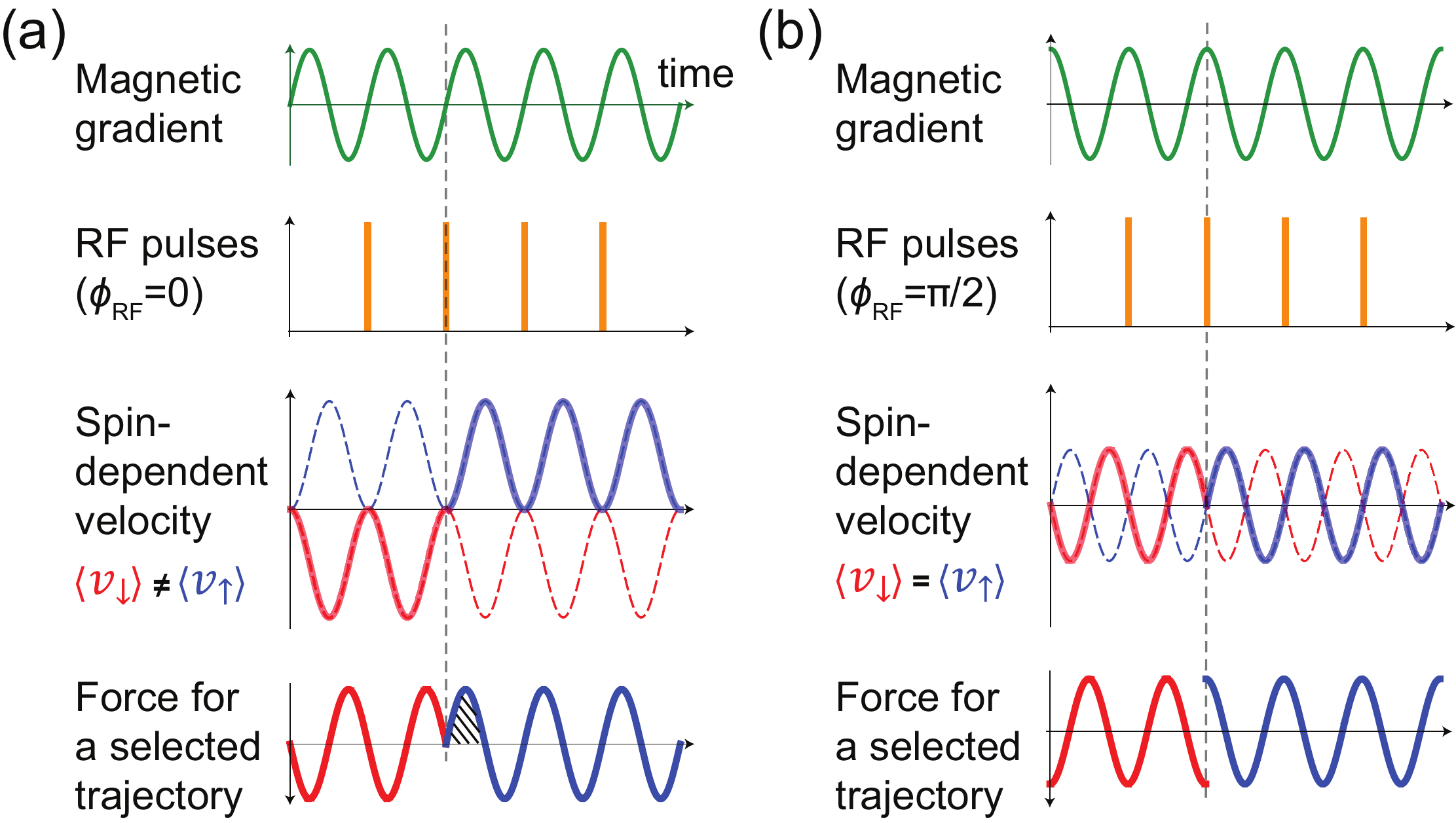}
\caption{\label{fig:schematic} Illustration of our scheme for creating a tunable atomic recoil momentum  with RF transitions using magnetic forces. (a) \& (b) shows the experimental conditions for $\phi_{RF}=0$ and $\phi_{RF}=\pi/2$, respectively. The spin-dependent forces and velocities are shown (as thick solid lines) for the amplitude of the wavefunction which is transferred from spin down (red) to up (blue) by the RF pulse marked by the gray dashed line. For $\phi_{RF}=0$, the average velocities $\langle v_{\downarrow}\rangle$ and $\langle v_{\uparrow}\rangle$ are different, which implies a finite recoil associated with the spin-flip. In contrast, $\langle v_{\downarrow}\rangle=\langle v_{\uparrow}\rangle$  for $\phi_{RF}=\pi/2$ and there is no recoil.}
\end{figure}

Figure \ref{fig:schematic} shows the time sequence of our scheme, which consists of a sinusoidal spin-dependent force $f(t)=g_F\mu_BB'_0\sin(\frac{2\pi}{T}t+\phi_{RF})\sigma_z$, where $g_F$ is the Lande factor, $\mu_B$ is the Bohr magneton and $B_0'$ is the magnitude of the magnetic gradient, and a synchronized sequence of short RF pulses at times $t=0, T, 2T ..$. The timing of the pulses with respect to the periodic force is described by the phase $\phi_{RF}$ which will determine the magnitude of the photon recoil. Each of the RF pulses couples the spin-up and spin-down states with the same velocity $v_{RF}$. For $\phi_{RF}=0$ the velocities averaged over a full cycle of the oscillating force, $\braket{v_\uparrow}$ and $\braket{v_\downarrow}$ are different. By flipping the spin, atoms experience an ``extra'' half-cycle of the magnetic acceleration (hatched area in Fig. \ref{fig:schematic}(a)), which transfer them to the state with a different averaged velocity, and, therefore, provides recoil. For the case $\phi_{RF}=\pi/2$, the time-averaged velocities for spin-up and spin-down are identical to $v_{RF}$. Therefore, an RF transition will not change the time-averaged velocities, and there is no recoil.

Using this semiclassical picture, we obtain for the amount of momentum transfer  $\hbar k=m(\langle{v}_{\uparrow}\rangle-\langle{v}_{\downarrow}\rangle)=\hbar k_0\cos\phi_{RF}$, where $k_0=\frac{g_F\mu_B}{\pi\hbar}B_0'T$. Next we discuss where the change in kinetic energy comes from. For an optical transition with recoil $\hbar k$ and an atom moving at initial velocity $v_{in}$, the resonance frequency is  shifted by the Doppler shift $k v_{in}$ and recoil shift $(\hbar k)^2/2m$ which ensures energy conservation.
However, in the current situation, energy can also come from the time-dependent magnetic force. Indeed, if we would apply a single RF $\pi$ pulse at phase $\phi_{RF}=0$, the time-averaged velocity would change by $\hbar k_0/m$, but the RF resonance frequency would be independent of velocity and $k_0$. However, if a series of RF pulses is used, as in Fig. \ref{fig:schematic}, the resonance is Doppler shifted and becomes velocity selective.  This can be seen by regarding the pulses as Ramsey pulses, and considering the phase evolution of the wavefunction between two pulses (see Supplement). The RF pulses create a superposition of spin up and spin down.
Between pulses, the phase evolution for spin up/down is solely determined by the kinetic energy $\alpha_{\uparrow\downarrow}= \frac{1}{\hbar}\int (mv_{\uparrow\downarrow}^2/2)dt$, leading to a phase difference $\delta \alpha =\frac{1}{\hbar}(m(\langle{v}_{\uparrow}\rangle-\langle{v}_{\downarrow}\rangle) v_{RF} ) T = kv_{RF} T$ after one period of shaking, where $v_{RF}=(v_{\uparrow}+v_{\downarrow})/2$ is the common velocity at the moment of RF pulse.
With $\langle{v}_{\downarrow}\rangle=v_{RF}-\hbar k /2m$ we find that for resonant excitation, the RF frequency has to compensate for this phase shift by the Doppler detuning $k \langle v_{\downarrow}\rangle$ and the recoil shift $(\hbar k)^2/2m$.

We can show more formally, that our scheme creates an RF photon with recoil.  Periodic Hamiltonians are often treated by Floquet theory \cite{Dalibard,Polkovnikov,Eckardt2015,Gediminas2017,Eckardt2017}, which provides an expression for an effective Hamiltonian $\hat{H}_{eff}$ describing the slow time evolution of the system averaged over the fast  micromotion with period $T$.  However, in the standard treatment the effective Hamiltonian is not unique and may depend on the initial time when the periodic drive is switched on. We adopt the approach of reference \cite{Dalibard} where the evolution of the quantum system with periodic drive is expressed by an effective Hamiltonian independent of initial and final times $t_i$, $t_f$ and a kick (micromotion) operator $\hat{K}$, which describes the initial kick due to a sudden switch on and the subsequent micromotion, shown as

\begin{equation}
\hat{U}(t_f,t_i)=e^{-i\hat{K}(t_f)}e^{-i\hat{H}_{eff}(t_f-t_i)}e^{i\hat{K}(t_i)}.
\end{equation}

For our scheme, the time-dependent Hamiltonian of the system in the frame rotating with the RF drive after the rotating-wave approximation is

\begin{multline}
\hat{H}=\frac{\hat{p}^2_z}{2m}+\hbar k_0\hat{z}\frac{\pi}{T}\sin(\frac{2\pi}{T}t+\phi_{RF})\hat{\sigma}_z\\
-\frac{1}{2}\hbar\delta_{RF}\hat{\sigma}_z+\hbar\Omega\hat{\sigma}_xT\sum\limits_n\delta(t-nT),
\end{multline}
where $\delta_{RF}$ is the RF detuning with respect to the atomic resonant frequency and $m$ is the atomic mass. The short RF pulses are represented as a series of delta-functions with effective Rabi frequency $\Omega$.

Through the derivation shown in the Supplement, we obtain
\begin{equation*}
\hat{H}_{eff}=
\begin{pmatrix}
\frac{\hat{p}^2_z}{2m}+\frac{1}{16}\frac{\hbar^2k_0^2}{m}-\frac{\hbar \delta_{RF}}{2} & \hbar \Omega e^{-ik_0\cos\phi_{RF}\cdot z}\\
\hbar \Omega e^{ik_0\cos\phi_{RF}\cdot z}  & \frac{\hat{p}^2_z}{2m}+\frac{1}{16}\frac{\hbar^2k_0^2}{m}+\frac{\hbar \delta_{RF}}{2}
\end{pmatrix}
\end{equation*}
\begin{equation}
\label{eq:HeffKick}
\hat{K}(t)=-ik_0z\hat{\sigma}_z\cos(\frac{2\pi}{T} t+\phi_{RF}).
\end{equation}
The effective Hamiltonian is identical to the one for a two level atom driven by a photon field at frequency $\omega_{RF}$ and with wavevector $k$, which confirms our discussion above about recoil momentum and Doppler shift. The term $\frac{1}{16}\frac{\hbar^2k_0^2}{m}$ is the kinetic energy due to micromotion.

We implemented this scheme using  a thermal cloud of approximately $1\times10^{5}$ \textsuperscript{23}Na atoms at 380 nK in a crossed optical dipole trap with trapping frequencies $(\omega_{x}, \omega_{y}, \omega_{z}) = 2\pi\times (98, 94, 25)$ Hz corresponding to Gaussian radii of $19.5\;\rm\mu m$, $20\;\rm\mu m$ and $68\;\rm\mu m$ respectively. The $\left|m_{F}=-1\right>$ and $\left|m_{F}=0\right>$ states of the $F=1$ hyperfine manifold of the atoms were used to form a pseudospin-1/2 system, which will be referred to as $\left|\uparrow\right>$ and $\left|\downarrow\right>$ states, respectively. The $\left|m_{F}=1\right>$ state was decoupled from this 2-level system through the quadratic Zeeman effect at a bias field of 11.4 Gauss. Since there is no micromotion in the ``non-magnetic" $\left|m_{F}=0\right>$ state, the maximum momentum transfer $\hbar k_0$ is reduced by a factor of two compared to the discussion above.

The oscillating magnetic force was created by a time-dependent 3D quadrupole field.  Along the bias field direction $z$, this provides a 1D periodic force. Orthogonal to the bias field, the periodic potential is quadratic --- there is no net force, only a (negligible) modulation of the confinement.  The amplitude of the magnetic field gradient was 48 G/cm at a frequency of 5 kHz, implying a recoil  $k_0= 0.07 k_L$ where $\hbar k_L$ is the recoil of the resonant transition at 589 nm, with a recoil velocity $ \frac{\hbar k_L}{m}=2.9$ cm/s. The field gradient was calibrated using Stern-Gerlach deflection during ballistic expansion of a Bose-Einstein condensate.  We also calibrated the recoil $k_0$ directly by measuring the momentum transfer to a cloud in the $\left|m_{F}=-1\right>$ state during half-cycle of the magnetic shaking. The two calibrations agreed to within the accuracy of measurement.

To resolve Doppler shifts of 200 Hz, sub-milliGauss stability was needed. Any asymmetry of the periodic magnetic field gradient leads to a time-averaged DC field gradient resulting in an inhomogeneous Zeeman shift which had to be suppressed at the 100 Hz level.  Finally, the applied magnetic fields were modified by eddy currents in the stainless steel chamber, which had to be accounted for (see Supplement).

The goal of the experimental demonstration was to show that the RF transition is now Doppler sensitive due to the recoil transfer. The spinflip transitions were driven by 4 $\rm{\mu}$s long RF pulses at 8 MHz with a Rabi frequency of 10 kHz resulting in approximately $\pi/12$ pulses and an average Rabi frequency of $\Omega= 200$ Hz.  Since it was not possible to switch off the shaking coils on micro-second time scales, the RF pulses had to be applied with the magnetic shaking present. However, by carefully synchronizing the pulses with the zero-crossing of the gradient ($\phi_{RF}=0$ or $\pi$) and suppressing any time-dependent bias field (by carefully aligning the dipole trap to the center of the quadrupole field) we suppressed spatial and temporal Zeeman shifts well enough to make the spin-flips uniform across the cloud (see Supplement).

The RF pulses and the shaking were applied while the atoms were trapped to ensure that the velocity distribution is independent of position.  In time-of-flight (TOF), this is no longer the case, and any residual Zeeman shift gradients could lead to velocity selection. To avoid broadening of the Doppler selected velocity groups by the trapping potential, the total interrogation was chosen to be 1.6 ms, much shorter than the trap period along the $z$ direction. This time is also comparable to the coherence time due to the ambient magnetic field stability. Based on these considerations, we applied a pulse sequence of 2 ms consisting of 10 magnetic shaking cycles with 9 RF pulses across them.

\begin{figure}[t]
\includegraphics[width = 8.08cm]{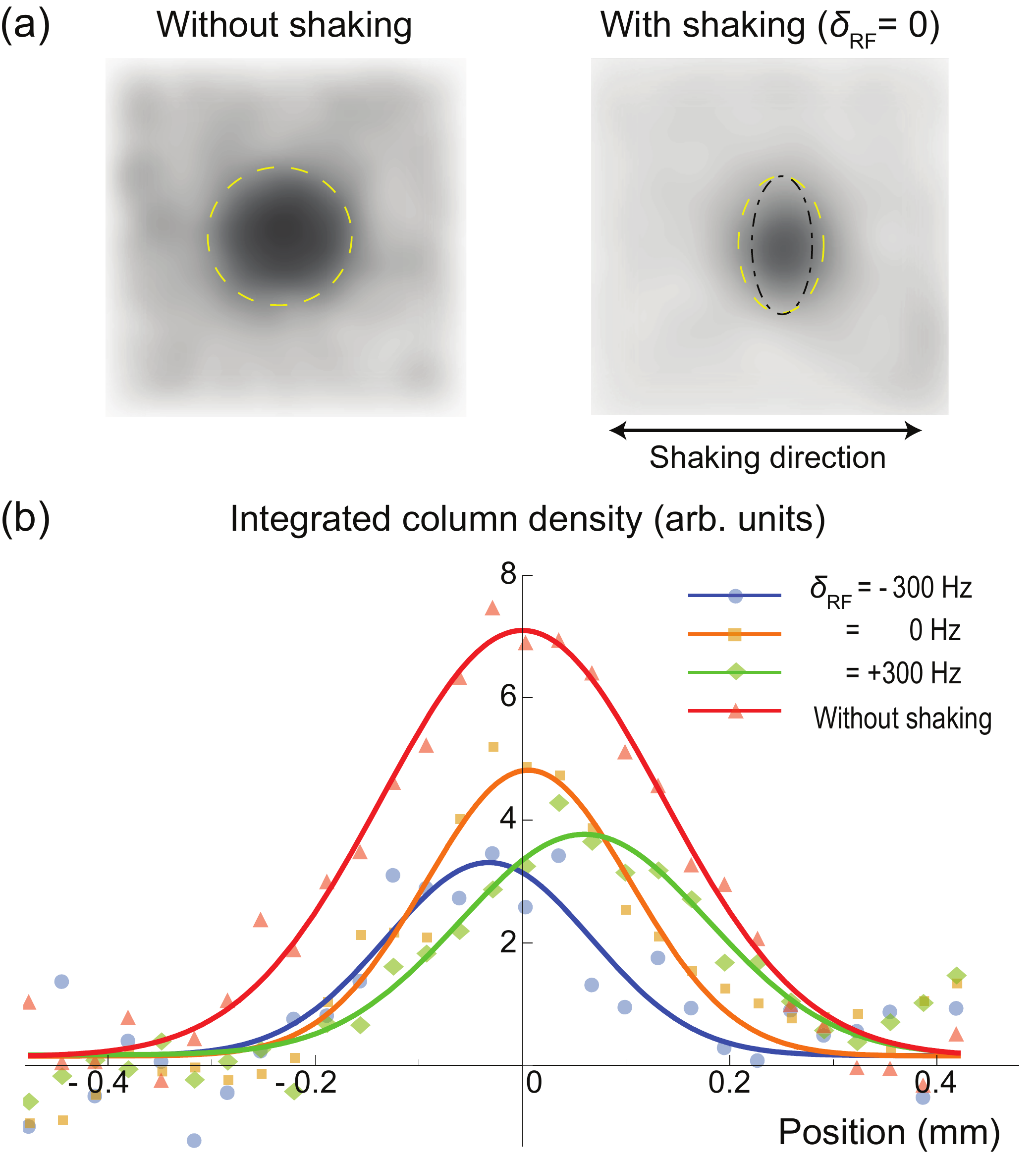}
\caption{\label{fig:Doppler} Observation of velocity selective RF transitions. (a) Absorption images of the spin-flipped atoms after 12 ms of TOF with and without magnetic shaking.  The yellow dashed ellipses have major and minor axes obtained as FWHM of Gaussian fits. After TOF, the thermal could expands by a factor of $2.13$, thus a single-velocity class is narrower than the thermal cloud by $1/2.13\approx0.47$. The Fourier limit of our velocity selection increases this to $0.50$, and inclusion of eddy currents further modifies it to $0.45$ (dashed-dotted line). The field of view is 1 mm by 1 mm. (b) Integrated column density distribution obtained from absorption images like those in (a), for different detunings of the RF frequency. The solids lines are Gaussian fits to the data points. The RF phase was at $\phi_{RF}$ = 0 to maximize Doppler sensitivity.}
\end{figure}

\begin{figure}[t]
\includegraphics[width = 8.34cm]{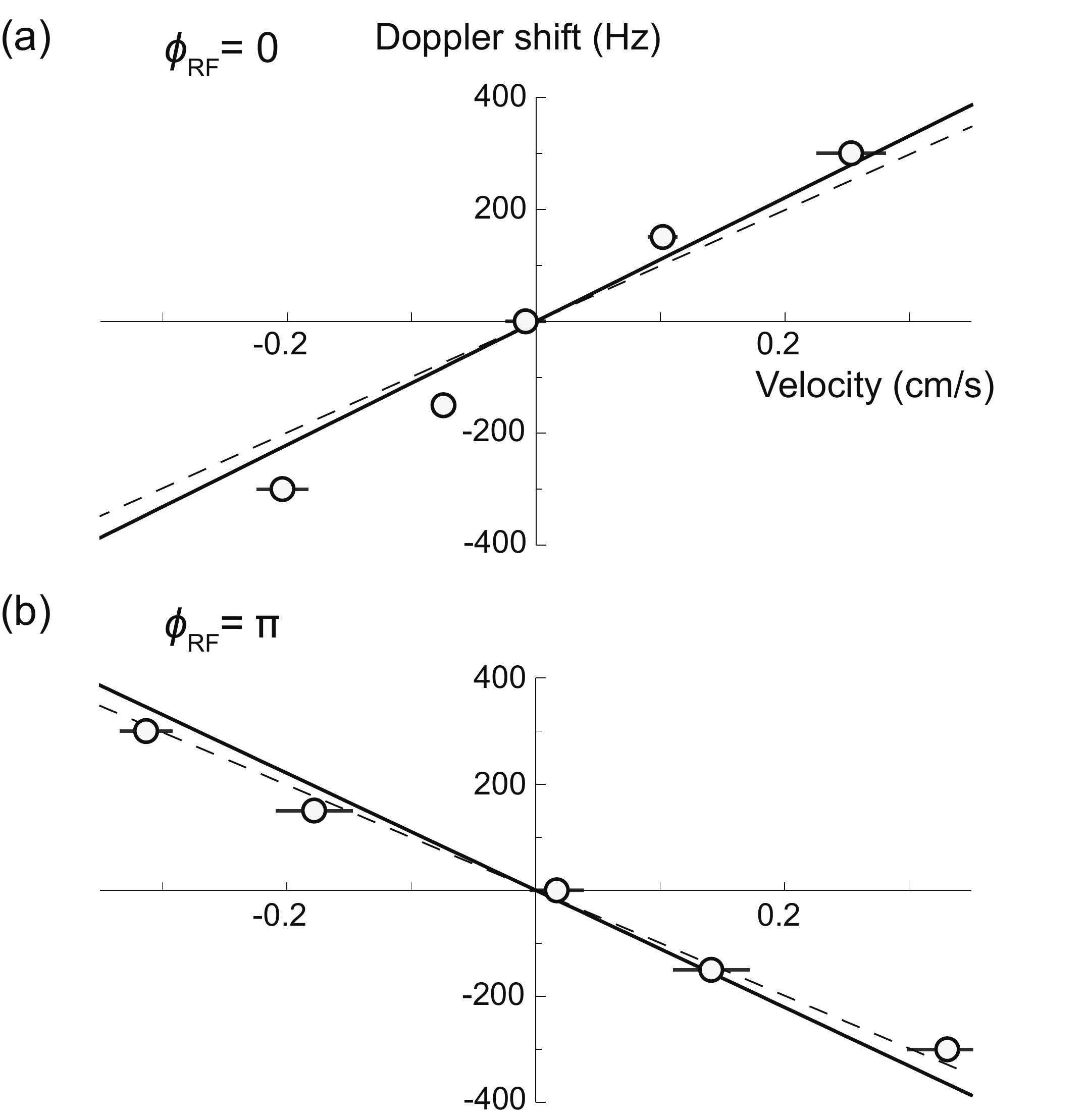}
\caption{\label{fig:rfphase} Observation of RF transitions with Doppler shifts. (a) \& (b) Central velocities of the spin-flipped atomic distribution (as in Fig. \ref{fig:Doppler}(b)) are shown as a function of RF detuning for $\phi_{RF}=0$ and $\phi_{RF}=\pi$, respectively. Shifting the RF phase changes the sign of the Doppler shift and therefore the direction of the recoil momentum.  The solid line represents the predicted Doppler shifts based on the calibration of recoil momentum. The dashed line takes into account the effects of eddy currents (see Supplement).  The error bars are  $1\sigma$.}
\end{figure}

The temperature of the cloud was chosen to be high enough that the Doppler width of 3 kHz (FWHM) was larger than our spectral resolution, mainly Fourier limited to 625 Hz by the 1.6 ms pulse sequence.  Due to the Doppler shift, different detunings of the RF selected different velocity groups which were observed in ballistic expansion (Fig. \ref{fig:Doppler}).  The width of the observed spin flipped slices is almost completely determined by the original spatial size of the cloud since the expansion time of $\tau= 12$ ms was only twice the inverse of $\omega_z$.  The TOF was limited by signal-to-noise, given the constraints discussed above for cloud temperature and trap frequencies. Fortunately, even for small TOF, the displacement of the center of the spinflipped atoms is exactly $v \tau$, which could be accurately measured as a function of RF detuning, as shown in Fig. \ref{fig:rfphase}.  The observed Doppler shift is in agreement with the theoretical treatment above and confirms that RF photons have been Floquet engineered to have recoil of $k=0.07 k_{L}$.

The dependence of the recoil on the RF phase was demonstrated by shifting the RF phase from 0 to $\pi$ (Fig. \ref{fig:rfphase}(b)). The Doppler shift and therefore the direction of the recoil changed sign. This observation confirmed that the selection of slices in Fig. \ref{fig:Doppler} is not due to time-averaged magnetic field gradients, which don't depend on the RF phase. We couldn't experimentally explore $\phi_{RF}=\pi/2$, since this would have required to pulse on the RF at the maximum field gradient which would have caused large spatially dependent detunings.

Our work realizes the key element of proposed schemes \cite{Ueda, Gediminas} for spin-orbit coupling of ultracold atoms with magnetic forces and without lasers.  The Hamiltonian (Eq. (\ref{eq:HeffKick})) which we have implemented is, by a unitary transformation, equivalent to a Hamiltonian with spin-dependent gauge fields \cite{DalibardGauge},

\begin{equation}
\hat{H}_{SOC}=\frac{1}{2m}(\hat{p}_z-\frac{1}{2}A\hat{\sigma}_z)^2+\hbar\Omega\hat{\sigma}_x-\frac{\hbar\delta_{RF}}{2}\hat{\sigma}_z.
\end{equation}
We note that reference \cite{Luo-supp} obtains the same Hamiltonian as stroboscopic Floquet Hamiltonian.
The gauge field $A=\hbar k_0\cos\phi_{RF}$ is equal to the recoil momentum transfer $\hbar k$ which depends on the RF phase $\phi_{RF}$.  Previous experimental studies claimed the realization of spin-orbit coupling and gauge fields purely by magnetic shaking, without RF transitions \cite{Luo2016, Wu2017}.  These claims are ambiguous based on our discussion here:  without RF coupling, the momentum transfer and the gauge field are not defined and can be transformed away with a gauge transformation.  According to Eqs. (1) and (3), pure magnetic shaking leads only to a kick operator for the micromotion, and the effective Hamiltonian is the free particle Hamiltonian. Therefore, all observations in Refs. \cite{Luo2016, Wu2017} are related to an initial kick and micromotion and not to a modified effective Hamiltonian.

\begin{figure}[t]
\includegraphics[width = 8.6cm]{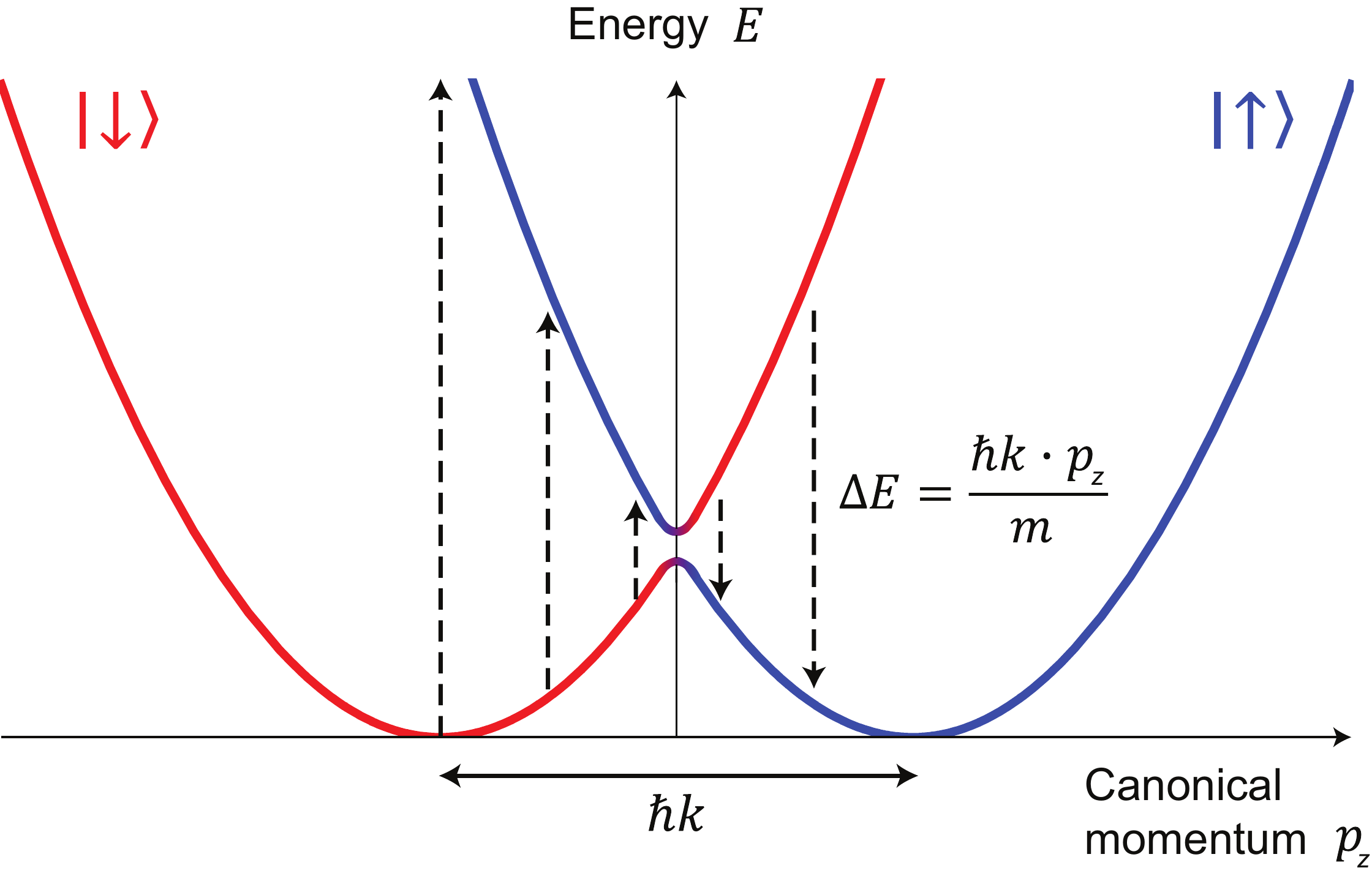}
\caption{(color)\label{fig:shakingSOC} Energy-momentum dispersion relations for spin-orbit coupled spin 1/2 states. The two minima are separated by the recoil momentum $\hbar k$.  The vertical dashed arrows show spinflip transitions. Their lengths are given by the Doppler and recoil shifts.}
\end{figure}

In the presence of gauge fields, there are two momenta: the mechanical or kinetic momentum $(p_z\pm\frac{1}{2} A)$, and the canonical momentum $p_z=mv_{RF}$.  In our scheme, they can both be directly observed and have a very transparent meaning:  the kinetic momenta are the time-averaged momenta $m \langle{v}_{\uparrow}\rangle$, $m \langle{v}_{\downarrow}\rangle$. The canonical momentum is the instantaneous momentum during the RF pulse. Using canonical momentum all couplings and transitions between the two spin states are vertical (Fig. \ref{fig:shakingSOC}). The dashed lines illustrate the transitions observed in our experiment. Away from the spin gap the energy separation is dominated by Doppler and recoil shifts.

The magnetic spin-orbit coupling scheme realized here completely eliminates heating from spontaneous emission which is a limiting factor for the two-photon Raman schemes \cite{Hui}.  However, as in any Floquet schemes, the micromotion can lead to heating. Although the spatial amplitude of the micromotion can be suppressed by faster modulation, the velocity amplitude is fixed as $\hbar k_0/2m$. The associated kinetic energy can be transferred to the secular motion by elastic collisions between the two spin states which occur at a rate $\frac{n}{2} \sigma_{\uparrow\downarrow} v_{rel}$, with total density $n$, inter-spin collision cross section $\sigma_{\uparrow\downarrow}$, and relative velocity $v_{rel}$ between the spin states. For a thermal cloud at a temperature $T$, the increase of energy $\dot{E}$ is $\dot{E} \propto n\sigma_{\uparrow\downarrow}{k_0}^2 \sqrt{T}$, while for a condensate, $\dot{E} \propto n\sigma_{\uparrow\downarrow}{k_0}^3$.  The same expression holds for degenerate Fermi gases with $k_0 \ll k_{\rm F}$, where elastic collisions are Pauli suppressed by a factor of $(k_0/k_F)^2$ (see Supplement). For a sodium condensate with $n \sim 10^{14}\;\rm cm^{-3}$, we observed a lifetime of $\sim 8\;\rm s$ at $k_0 = 0.05k_{\rm L}$, which is much longer than the inverse of mean-field interaction time. Therefore, it should be possible to study interactions in spin-orbit coupled systems \cite{Stringari}.

There are possible extensions of magnetic spin-orbit coupling scheme.  One is to use the TOP trap configuration \cite{Cornell} where a constant gradient is combined with a rotating bias field in the $x$-$y$ plane which creates a rotating force.  A sequence of RF pulses generates 1D spin-orbit coupling with recoil $k$ along the $\cos\phi_{RF}\mathbf{e}_x+\sin\phi_{RF}\mathbf{e}_y$ direction. The RF phase now controls the direction of the recoil. With this scheme, it should be possible to create larger recoils $k$, since it is easier to create stationary magnetic field gradients than rapidly oscillating ones. Magnetic shaking can realize two-dimensional spin-orbit coupling \cite{Gediminas, Ueda}.  In this case, fast switching of the bias field direction is required in order to project the spin states rather than letting the spins adiabatically follow. For this, it will be beneficial to use small coils or wires on an atom chip, and not large coils of 10 cm size as in our work.

In conclusion, we demonstrated how magnetic shaking can be used to endow an RF photon with large and tunable recoil and realized the basic element of spin-orbit coupling without lasers and therefore without heating by spontaneous light scattering.  This scheme can be applied to any atom or molecule with non-zero  spin in the ground state, and is independent of the structure of electronically excited states.

\begin{acknowledgments}
We would like to acknowledge Will Lunden for experimental assistance, Ivana Dimitrova for critical reading of the manuscript, and Viktor Novi\v{c}enko for discussions.
We acknowledge support from the NSF through the Center for Ultracold Atoms and award 1506369, from ARO-MURI Non-equilibrium Many-body Dynamics (Grant No. W911NF-14-1-0003), from AFOSR-MURI Quantum Phases of Matter (Grant No. FA9550-14-1-0035), from ONR (Grant No. N00014-17-1-2253) and a Vannevar-Bush Faculty Fellowship.
Part of this work was performed at the Aspen Center for Physics, which is supported by NSF grant PHY-1607611.
\end{acknowledgments}

\newpage
\onecolumngrid

\appendix

\setcounter{equation}{0}
\setcounter{figure}{0}
\makeatletter
\renewcommand{\theequation}{S\arabic{equation}}
\renewcommand{\thefigure}{S\arabic{figure}}
\renewcommand{\bibnumfmt}[1]{[S#1]}
\renewcommand{\citenumfont}[1]{S#1}

\section{Supplemental Material for\\ ``How to dress radio-frequency photons with tunable momentum''}

\section{Experimental implementation of magnetic shaking}

To realize magnetic shaking, we drove a sinusoidal current through a pair of anti-Helmholtz coils along the $x$-axis while there was a fixed bias field of 11.4 G aligned to the $z$-axis. The sinusoidal current was provided by a DC power supply and four insulated gate bipolar transistors connected in a H-bridge configuration. The transistors created a square wave voltage modulation, which resulted in a sinusoidal current due to the frequency response of the coils. A capacitor was connected in series to eliminate the imaginary component of the impedance coming from the inductance of the coils. The amplitudes of the current and the voltage were $70 \;\rm A$ and $70 \;\rm V$. The real impedance of 1 $\Omega$ is mainly due to eddy currents in the stainless steel vacuum chamber and is much larger than the resistance of the coils (0.1 $\Omega$). The combined fields result in a periodic 1D magnetic force along the $z$-axis $B_{z}'(t)=B_{0}'\mathrm{sin} (\omega t + \phi_{RF})$, where $B_{0}'$ = 48 G/cm, $\omega=2\pi\times$ 5 kHz, and $\phi_{RF}$ determined from the relative phase between the magnetic gradient modulation and the radio frequency (RF) pulses. Larger recoil momentums can be realized by either increasing $B_{0}'$ or decreasing $\omega$.

\section{Adjustments to the magnetic field profile}
The observation of Doppler shifts at the 200 Hz level required careful control of magnetic Zeeman shifts.  Three critical adjustments were done.

(1)\emph{Symmetry of the modulated magnetic field gradient}: If inhomogeneous Zeeman shifts across the cloud are comparable or larger than Doppler shifts, the spinflips are no longer velocity selective since there is always a local Zeeman shift to compensate for the Doppler shift.  Therefore, the magnetic field gradient averaged over one modulation cycle $\langle B'\rangle$, had to be zeroed: $g_F \mu_B \langle B'\rangle D \ll k v$, where $D$ is the length of the cloud. To avoid transient asymmetries from the turn-on process of the periodic magnetic gradient, we added a pre-shaking period of 3 ms before the spectroscopic sequence.  This didn't affect the trapped atom cloud, since the atoms were initially in the non-magnetic  $\left|m_{F}=0\right>$ state.  After the pre-shaking, we achieve $\langle B'\rangle \approx$ 20 mG/cm, implying a time-averaged differential Zeeman shift across the cloud of less than 100 Hz. $\langle B'\rangle$ was determined from converting the measurement of time-averaged current asymmetry to the time-averaged magnetic gradient asymmetry using the Stern-Gerlach calibration. As a final check, we added asymmetries on either the positive or negative side of the sinusoidal current to create $\langle B'\rangle \approx \pm$ 100 mG/cm, and for both cases observed a slight increase in the width of the velocity-selected atom slice confirming that the residual asymmetry of the magnetic gradient modulation was negligibly small.
\vspace{3mm}
\newline
The following two adjustments addressed the issue that the RF pulses were not delta functions, but had a duration of 4 $\mu$s. The presence of Zeeman shifts comparable or larger than the Fourier width of a single pulse would reduce the RF pulse area. For our parameters, a 45 kHz detuning will reduce the pulse area by 5 percent (and therefore the single pulse excitation probability by 10 percent).

(2)\emph{Minimize modulation of magnetic bias field}: The time-dependent gradient creates also a time-dependent bias field given by the gradient times the displacement of the atoms from the origin of the magnetic quadrupole field. 60 $\mu$m away from the origin, the bias field changes by 30 mG during the 4 $\mu$s RF pulse. To minimize the reduction of the RF pulse area, the optical trap was aligned with the center of the quadrupole field to within 1 $\mu$m. This was done by minimizing the shift in the RF resonant frequency when a stationary gradient field was added to the constant magnetic bias field. In addition, the eddy currents created a time-dependent bias field, which was compensated by RF detuning. The detuning and the timing of the RF pulses (described below) were adjusted together in order to maximize the fraction of spin-flipped atoms.

(3)\emph{Timing of the RF pulses with respect to the magnetic modulation}: The goal was to pulse on the RF while the magnetic field gradient crosses zero.  A 5 $\mu$s offset would imply a gradient of 7.5 G/cm and a differential magnetic field along the cloud of 50 mG.  In the presence of strong gradients, the short RF pulse is resonant only for a small part of the cloud.  Therefore, we could find the optimum condition by scanning both the timing and the detuning of the RF pulses until the measured total fraction of the spin-flipped atoms is maximized. The optimum time was offset by 2 $\mu$s from the zero-crossing of the current through the gradient coils, possibly due to eddy currents.

To summarize, we were optimizing three parameters, which are trap position, timing of the RF pulse, and RF detuning. The optimal position minimizes temporal variation of the bias field, optimal timing of RF minimizes B' during the pulse, and optimal detuning compensates for any bias field at the time of the pulse.

\section{Effects of Induced eddy Currents}

The modulated magnetic field gradient $B'(t) = B_0'\sin{\omega t}$ induced eddy currents in the stainless steel vacuum chamber. From our observations, we inferred that the main effect was caused by an induced oscillating bias field $\vec{B}_{\rm ec}(t) = B_{\rm ec}\sin{(\omega t+\phi)}\textbf{e}_y$ along $y$ with the same modulation frequency $\omega$ and a relative phase delay $\phi$. This oscillating bias field led to a $y$-component of the oscillating force. As a result, the effective recoil and velocity selection are tilted away from the $z$ direction, and the selected velocity slices are rotated in the $y-z$ plane.

\begin{figure}[h]
\includegraphics[width = 13cm]{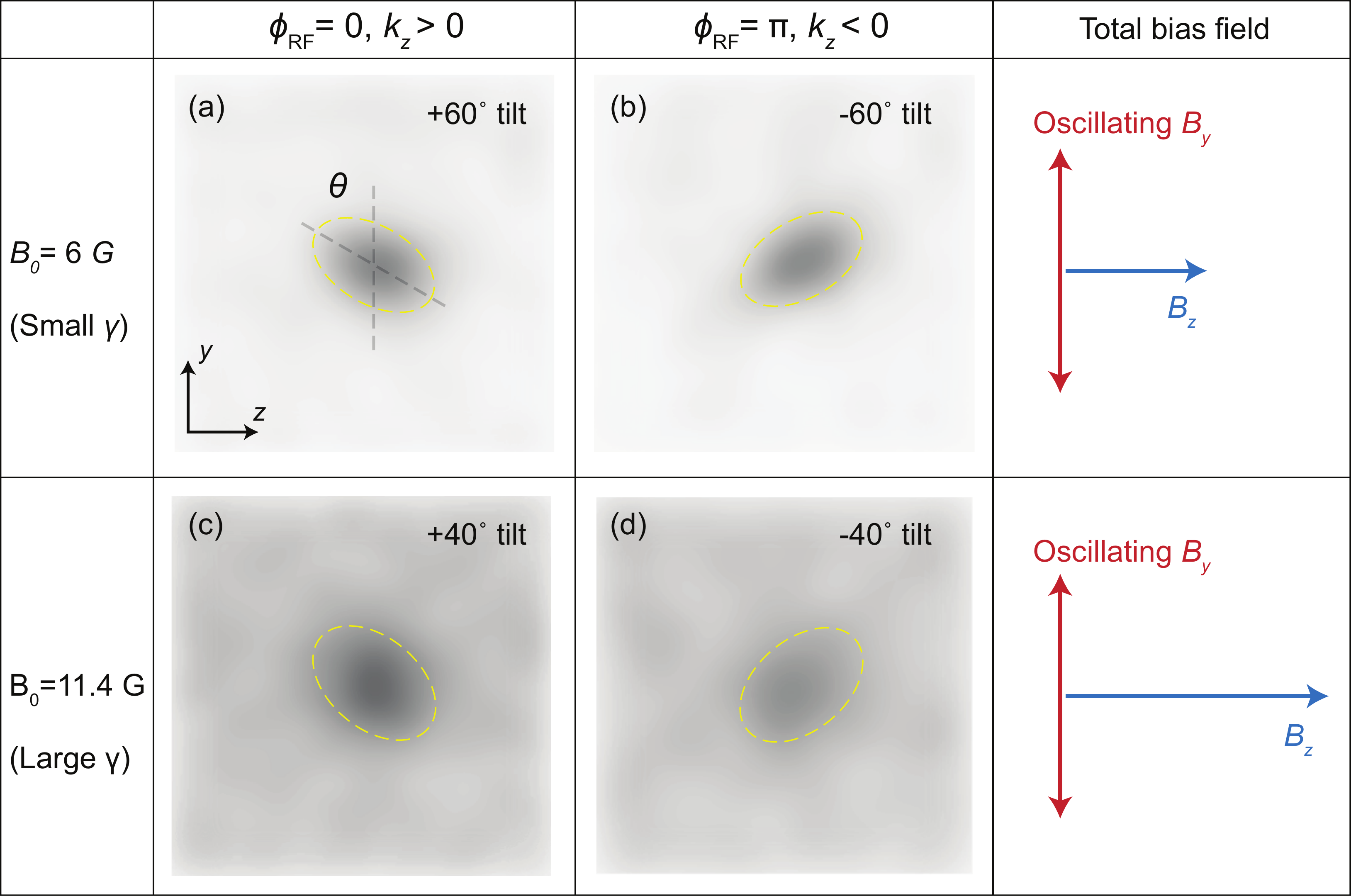}
\caption{\label{fig:rotation}Effect of eddy currents on observed velocity-selected atom slices. The induced bias field along $\textbf{e}_y$ led to a $y$-component of the oscillating force, resulting in velocity selectivity in $\textbf{e}_y$ and therefore tilting of the resonant velocity slice in the $y-z$ plane. The tilt angle depends on the static bias field $B_0$ and the RF phase $\phi_{RF}$. The dashed lines are guides to the eye. }
\end{figure}

In a simplified model, the total magnetic field experienced by the atoms is
\begin{equation}
\vec{B} = \left[ B_0+B_0'\sin{(\omega t)}z\right]\textbf{e}_z +\left[B_{\rm ec}\sin{(\omega t+\phi)}+ B_0'\sin{(\omega t)}y\right]\textbf{e}_y-2B_0'\sin{(\omega t)}x\textbf{e}_x,
\end{equation}
with a magnetic field strength
\begin{equation}
\begin{split}
|\vec{B}|&=\sqrt{\left[ B_0+B'_0\sin{(\omega t)}z\right]^2 +\left[B_{\rm ec}\sin{(\omega t+\phi)}+ B'_0\sin{(\omega t)}y\right]^2+\left[2B_0'\sin{(\omega t)}x\right]^2}\\
&\approx B_0\sqrt{1+\gamma^2\sin^2{(\omega t+\phi)}}+\frac{B_0'z+\gamma\sin{(\omega t+\phi)}B_0'y}{\sqrt{1+\gamma^2\sin^2{(\omega t+\phi)}}}\sin{(\omega t)}\\
\end{split}
\end{equation}
here $\gamma = B_{\rm ec}/B_0$. The first term corresponds to a time varying homogeneous bias field resulting in a \textit{velocity-independent} effective detuning of the RF transition. The oscillating magnetic field gradients along the $z$ and $y$ directions are
\begin{equation}
\begin{split}
&\frac{\partial |\vec{B}|}{\partial z} = \frac{B_0'}{\sqrt{1+\gamma^2\sin^2{(\omega t+\phi)}}}\sin{(\omega t)},\\
&\frac{\partial |\vec{B}|}{\partial y}=\frac{\gamma B_0'}{\sqrt{1+\gamma^2\sin^2{(\omega t+\phi)}}}\sin{(\omega t)}\sin{(\omega t+\phi)}.\\
\end{split}
\end{equation}
It should be noted that the gradient in $\textbf{e}_y$ oscillates at $2\omega$, twice the frequency of the driving.

The phase delay $\phi$ is determined by the magnetic properties of the vacuum chamber. We modeled the chamber as a LC circuit with a self inductance $L_{\rm Ch}$ and a resistance $R_{\rm Ch}$, and obtain $\phi = \arctan{(\omega L_{\rm Ch}/R_{\rm Ch})}+\pi/2$. Our observations imply $R_{\rm Ch}\gg \omega L$ , $\phi \approx \pi/2$, resulting in an effective recoil component in the in $y$ direction with
\begin{equation}
\label{eq:k2d}
\begin{split}
k_{{\rm so},y} &= \frac{1}{T}\int_0^{T}\left(\int_0^t \frac{\gamma B_0'}{\sqrt{1+\gamma^2\cos^2{(\omega t')}}}\sin{(\omega t')}\cos{(\omega t')}\;{\rm{d}}t'\right){\rm{d}}t\\
&= \frac{1}{T}\int_0^{T}\left(\int_0^t \frac{\gamma B_0'}{2\sqrt{1+\gamma^2\cos^2{(\omega t')}}}\sin{(2\omega t')}\;{\rm{d}}t'\right){\rm{d}}t.\\
\end{split}
\end{equation}
Consequentially, the Doppler shift is modified as
\begin{equation}
\label{eq:res}
\delta\omega = k_{y}v_y+k_{z}v_z,
\end{equation}
directly observed as a rotation of the velocity slice with an angle  $\theta = \arctan{(k_{y}/k_{z})}$ in the time-of-flight images, as shown in Fig. \ref{fig:rotation}.

We verified two predictions of this model : the angle $\theta$ of the rotation decreased with stronger static bias field $\vec{B}_0$ which lowered $\gamma$ (Fig. \ref{fig:rotation}(a). and Fig. \ref{fig:rotation} (c)). Due to the $2\omega$ oscillating frequency of the $y$ force, $k_{y}$ did not change sign when the RF phase $\phi_{RF}$ was shifted from 0 to $\pi$ in contrast to $k_z$, and therefore the rotation angle flipped from $\theta$ to $-\theta$, as suggested by Eq. (\ref{eq:res}) and shown in Fig. \ref{fig:rotation}.

In the future, the effects of the induced eddy current can be suppressed by using an even stronger static bias field $\vec{B}_0$ or by conducting the experiment in a glass cell.

What we have described so far applies to free space or to an isotropic trap. However, the optical trap in the experiment is anisotropic. For zero time-of-flight, in the $y-z$ plane, the minor axis of the ellipsoidal cloud is oriented along $y$, $\theta = \pi/2$. For long time-of-flight, the angle is solely determined by the velocity selection $\theta = \arctan{(k_{y}/k_{z})}$. For intermediate time-of-flight, as used in the experiment, the observed angle interpolates between these values. We calculate that the observed tilt angles of $60^\circ$ and $40^\circ$ (Fig. \ref{fig:rotation}) correspond to tilt angles of the bias field $\arctan{(k_{y}/k_{z})}$ of $53^\circ$ and $32^\circ$, respectively.

The observed tilt angles were used to infer the induced eddy currents. Equation (\ref{eq:k2d}) provided the dashed line for the predicted recoil $k$ in Figure \ref{fig:rfphase} of the main text.

\section{Bloch sphere representation of magnetic shaking and RF pulses}
\begin{figure}[h]
\includegraphics[width = 13cm]{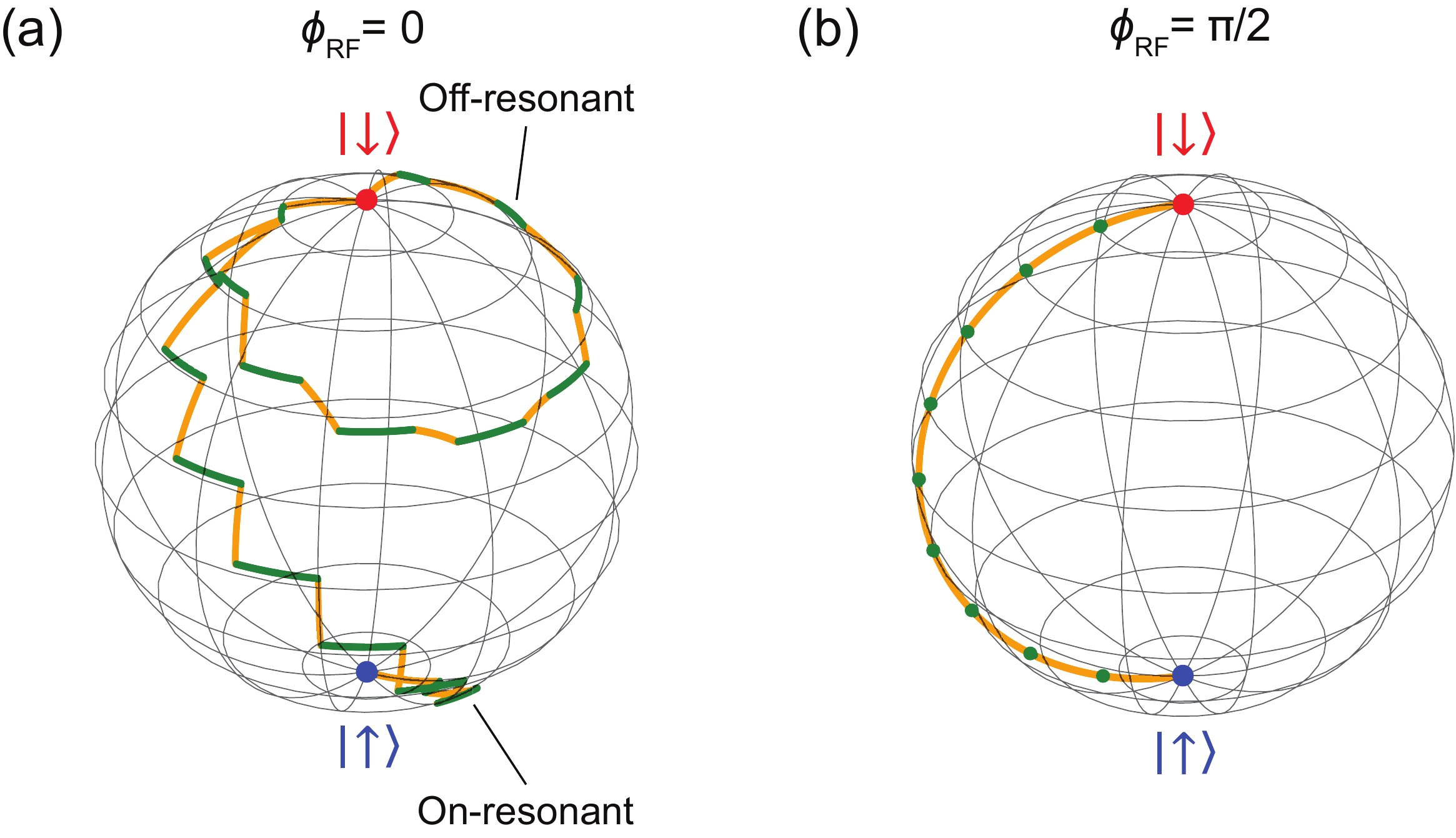}
\caption{\label{fig:bloch} Bloch sphere representation of magnetic shaking and RF pulses. (a) \& (b) Trajectories on the Bloch sphere for several periods of magnetic shaking (green solid lines representing $\alpha$ from the main text) and RF pulses (yellow solid lines representing $\beta$) for $\phi_{RF}=0$ and $\phi_{RF}=\pi/2$, respectively. Fig. (a) shows the trajectories for atoms with a finite initial velocity when the RF frequency is at $\omega_0$, the atomic resonance, and when it is detuned by the Doppler and recoil shift.   In (b), the RF frequency is at $\omega_0$,  the trajectory is independent of the atomic velocity, and there is no net rotation around the z-axis during a magnetic shaking cycle. The red (blue) dot represents the initial (final) spin state.}
\end{figure}

The evolution of the quantum system under magnetic shaking and RF pulses can be visualized using the Bloch sphere (Fig. \ref{fig:bloch}). In the frame rotating at the atomic RF resonance frequency $\omega_0$, each RF pulse of area $\beta$ rotates the Bloch vector around the $y$-axis by an angle $\beta$. In the absence of magnetic shaking, subsequent pulses would continue the rotation all the way down to the south  pole of the Bloch sphere and up again, resulting in Rabi oscillations at a rate $\beta/(2 \pi T)$.  However, due to the phase evolution discussed in the main text, the Bloch vector rotates around the z axis by an angle $\delta \alpha$, and therefore, the following RF pulse increase the polar angle by less than $\beta$.  After several cycles, the Bloch vector returns to the north pole without having ever reached the south pole, realizing off-resonant Rabi oscillations (Fig. \ref{fig:bloch}(a)).  However, if the RF frequency is shifted by the Doppler and recoil shift, the Bloch vector reaches the south pole again.  In contrast, for the phase $\phi_{RF}=\pi/2$, kinetic energies of the coupled spin up and down states are the same, irrespective of velocity, and therefore all atoms perform resonant Rabi oscillations (Fig. \ref{fig:bloch}(b)).  It should be noted that the evolution of the atomic wavefunction is the same if the RF frequency is detuned by an integer multiple of $\frac{2\pi}{T}$, similar to the situation in Ramsey spectroscopy.

\section{Micromotion Heating}
The fast micromotion can lead to heating due to elastic collisions between atoms in two spin states. An upper limit of the heating rate for a equal spin mixture can be estimated with the time-averaged kinetic energy $E_{\rm micro}$ of the micromotion and the inter-spin two-body collision rate $\Gamma_{\uparrow\downarrow}$. We obtain:
\begin{equation}
\Gamma_{\rm heating} \approx \frac{1}{2.7}\Gamma_{\uparrow\downarrow} E_{\rm micro}
\end{equation}
Here $\Gamma_{\uparrow\downarrow}= \frac{n}{2}\sigma_{\uparrow\downarrow} \overline{|v_{\uparrow\downarrow}|}$ where $\overline{|v_{\uparrow\downarrow}|}$ is the relative speed of the atoms in two spin states averaged over the ensemble, $n$ is the total density, and $\sigma_{\uparrow\downarrow}$ is the inter-spin \textit{s}-wave scattering cross section. The factor $2.7$ is the number of collisions required to distribute the energy to all three dimensions \cite{monroe1993}. $E_{\rm micro}$ is the kinetic energy of the relative micromotion between the spin up and down atoms, $(\hbar k_0)^2/4m$.

For a thermal cloud at temperature $T$, the velocity distribution is a Boltzmann distribution, and we estimate the heating rate to be
\begin{equation}
\begin{split}
\Gamma_{\rm heating}^{\rm thermal} &\approx \frac{n\sigma_{\uparrow\downarrow}}{2.7}\int {\rm d}v_{\uparrow}\int {\rm d}v_{\downarrow}\;|v_{\uparrow}-v_{\downarrow}|e^{-\frac{mv^2_{\uparrow}}{2kT}}e^{-\frac{mv^2_{\downarrow}}{2kT}}\frac{(\hbar k_0)^2}{8m}\\
&\approx \frac{n\sigma_{\uparrow\downarrow}}{12 m}\sqrt{\frac{kT}{\pi m}}(\hbar k_0)^2\\
\end{split}
\end{equation}
which gives the relative heating rate $\dot{T}/T \propto n\sigma_{\uparrow\downarrow}(\hbar k_0)^2/\sqrt{T}$. For our experiment conditions, we estimated $\dot{T}/T \approx 0.01 /\rm s$.

A Bose-Einstein condensate has negligible thermal velocity. The relative motion is dominated by the micromotion with $\overline{|v_{\uparrow\downarrow}|} \approx  \hbar k_0/m$. The heating rate therefore reads
\begin{equation}
\Gamma_{\rm heating}^{\rm BEC}\approx \frac{ n\sigma_{\uparrow\downarrow}}{9\pi m^2} (\hbar k_0)^3,
\end{equation}
where the numerical pre-factor is the result of the time-average. For a condensate with $n \sim 10^{14}\;\rm cm^{-3}$ at $k = 0.05k_{\rm L}$, we obtain $\Gamma_{\rm heating}^{\rm BEC} \approx h \times 90 \;\rm Hz/s$.

For a mixture of degenerated Fermi gases, Pauli blocking prevents the atoms to be scattered to already occupied states. For $(T/T_F)^2\ll (k_0/k_F)^2\ll1$, only atoms on the Fermi surface collide, resulting in a heating rate:
\begin{equation}
\begin{split}
\Gamma_{\rm heating}^{\rm Fermi} &\approx \frac{1}{2.7}\left(\frac{k_0}{k_{\rm F}}\right)^2\frac{n\sigma_{\uparrow\downarrow}2\hbar k_F}{m^2}\frac{(\hbar k_0)^2}{8}
\end{split}
\end{equation}
which is Pauli suppressed by a factor $(k_0/k_F)^2$.

\section{Derivation of Effective Hamiltonian}

In our scheme, the Hamiltonian in the frame rotating with the RF drive
is

\begin{equation}
\hat{H}(t)=\frac{\hat{p}_{z}^{2}}{2m}-\frac{1}{2}\delta_{RF}\hat{\sigma}_{z}+\frac{1}{2}\omega k_{0}\hat{z}\sin\left(\frac{2\pi}{T}t+\phi_{RF}\right)\hat{\sigma}_{z}+\Omega\hat{\sigma}_{x}T\sum_{n}\delta\left(t-nT\right)\,,\label{eq:H - Supplement}
\end{equation}
where we have applied the rotating-wave approximation and set $\hbar=1$.
To deal with the dynamics of such a periodically driven system we
shall apply two alternative approaches described below.

To eliminate the spin-dependent potential slope featured in the Hamiltonian
(\ref{eq:H - Supplement}), we go to the spin-dependent co-moving
frame via a time-dependent unitary transformation to the new state-vector
$\tilde{\left|\psi(t)\right\rangle} =\hat{R}_{z}^{\dagger}\left(t\right)\left|\psi(t)\right\rangle $,
similar to the one used in refs. \cite{SGediminas,SLuo-supp}:

\begin{equation}
\hat{R}_{z}\left(t\right)=\exp\left[-ik_{0}z\gamma\left(t\right)\hat{\sigma}_{z}/2\right]\,,\qquad\gamma\left(t\right)=\omega\intop_{0}^{t}\sin\left(\frac{2\pi}{T}t^{\prime}+\phi_{RF}\right)dt^{\prime}-C=-\cos\left(\frac{2\pi}{T}t+\phi_{RF}\right)\,,\label{eq:R_z - Supplement}
\end{equation}
where the integration constant $C$ entering $\gamma\left(t\right)$
has been taken to be $C=\cos\phi_{RF}$, so that $\gamma\left(t\right)$
averages to zero over a period. The reason of such a choice will be
discussed later on.

At the RF pulses where $t=nT$ the transformation $\hat{R}_{z}\left(nT\right)=\exp\left[ik_{0}z\cos\phi_{RF}\hat{\sigma}_{z}/2\right]$
describes a spin rotation by an angle $k_{0}z\cos\phi_{RF}$ around
the $z$ axis. As a result, the transformed Hamiltonian $\hat{\tilde{H}}\left(t\right)=\hat{R}_{z}^{\dagger}\left(t\right)\hat{H}\hat{R}_{z}\left(t\right)-i\hat{R}_{z}^{\dagger}\left(t\right)\partial_{t}\hat{R}_{z}\left(t\right)$
takes the form

\begin{equation}
\hat{\tilde{H}}\left(t\right)=\frac{1}{2m}\left(\hat{p}_{z}-\frac{1}{2}k_{0}\gamma\left(t\right)\hat{\sigma}_{z}\right)^{2}-\frac{1}{2}\delta_{\mathrm{RF}}\hat{\sigma}_{z}+\Omega\left[\cos\left(k_{0}z\cos\phi_{RF}\right)\hat{\sigma}_{x}+\sin\left(k_{0}z\cos\phi_{RF}\right)\hat{\sigma}_{y}\right]T\sum_{n}\delta\left(t-nT\right)\,.\label{eq:H-tilde - Supplement}
\end{equation}
 Note that unlike the spin-dependent potential gradient featured
in the original Hamiltonian (\ref{eq:H - Supplement}), the oscillating
momentum shift term $k_{0}\gamma\left(t\right)\hat{\sigma}_{z}$/2
is no longer proportional to the driving frequency and hence can be
considered as a small perturbation in the limit of high frequency
driving where $k_{0}\gamma\left(t\right)\ll\omega$ and also $\Omega\ll\omega$.
In that case it is appropriate to describe the evolution of the system
in terms of the zero-order effective Hamiltonian obtained by time
averaging of $\hat{\tilde{H}}\left(t\right)$ over a single driving
period, i.e. by the zero frequency component of the Hamiltonian $\hat{\tilde{H}}\left(t\right)$,
giving
\begin{equation}
\hat{H}_{eff}=\frac{\hat{p}_{z}^{2}}{2m}-\frac{1}{2}\delta_{RF}\hat{\sigma}_{z}+\Omega\cos\left(k_{0}z\cos\phi_{RF}\right)\hat{\sigma}_{x}+\Omega\sin\left(k_{0}z\cos\phi_{RF}\right)\hat{\sigma}_{y}+\frac{1}{16}\frac{k_{0}^{2}}{m}\,,\label{eq:H_eff - suppl}
\end{equation}
where the momentum shift has averaged to zero. The effective Hamiltonian
can be represented in a matrix form as:
\begin{equation}
\hat{H}_{eff}=\left(\begin{array}{cc}
\frac{\hat{p}_{z}^{2}}{2m}+\frac{1}{16}\frac{k_{0}^{2}}{m}-\frac{1}{2}\delta_{RF} & \Omega e^{-ik_{0}z\cos\phi_{RF}}\\
\Omega e^{ik_{0}z\cos\phi_{RF}} & \frac{\hat{p}_{z}^{2}}{2m}+\frac{1}{16}\frac{k_{0}^{2}}{m}+\frac{1}{2}\delta_{RF}
\end{array}\right)\,.\label{eq:H_eff matrix - suppl}
\end{equation}

The full dynamics includes also the micromotion. In the present situation
there are two origins of the micromotion. The first kind
comes from the time-dependence of the transformed Hamiltonian $\hat{\tilde{H}}\left(t\right)$.
However, in the limit of the large driving frequency this kind of
micromotion is negligibly small compared to the second type of micromotion
emerging due to the time-dependence of the unitary transformation
$\hat{R}_{z}\left(t\right)$. In fact, returning to the original representation
$\left|\psi(t)\right\rangle =\hat{R}_{z}\left(t\right)\tilde{\left|\psi(t)\right\rangle }$,
one arrives at the following time-evolution of the state-vector from
the initial to the final time
\begin{equation}
\left|\psi(t_2)\right\rangle =\hat{U}\left(t_2,t_1\right)\left|\psi(t_1)\right\rangle \,,\quad\mathrm{where}\quad\hat{U}\left(t_{2},t_{1}\right)=\hat{R}_{z}\left(t_{2}\right)e^{-i\hat{H}_{eff}\left(t_{2}-t_{1}\right)}\hat{R}_{z}^{\dagger}\left(t_{1}\right),\label{eq:evolution - Suppl}
\end{equation}
where the unitary transformation $\hat{R}_{z}\left(t\right)$ represents
a micro-motion operator.

The time evolution operator can be rewritten as

\begin{equation}
\hat{U}\left(t_{2},t_{1}\right)=e^{-i\hat{K}\left(t_{2}\right)}e^{-i\hat{H}_{eff}\left(t_{2}-t_{1}\right)}e^{i\hat{K}\left(t_{1}\right)}\,,\label{eq:evolution - Suppl-1}
\end{equation}
where
\begin{equation}
\hat{K}\left(t\right)=-k_{0}z\hat{\sigma}_{z}\cos\left(\frac{2\pi}{T}t+\phi_{RF}\right)\label{eq:K- Suppl}
\end{equation}
is a Hermitian micromotion (kick) operator. The choice of the integration
constant $C=\cos\phi_{RF}$ in the unitary transformation (\ref{eq:R_z - Supplement})
ensures that the micromotion operator $\hat{K}\left(t\right)$ averages
to zero over the driving period. Thus, the effective Hamiltonian and the
micromotion operators are defined in a unique way through the condition
$C=\cos\phi_{RF}$.

We now rederive the same effective Hamiltonian and micromotion operator using a rigorous high-frequency $1/\omega$ expansion. Appendix K of the ref. \cite{SDalibard} discusses Hamiltonians of a general form
\begin{equation}
\hat{H}(t)=\hat{H}_0+\hat{A}f(t)+\omega \hat{B}g(t)
\end{equation}
and derives expansions for an effective Hamiltonian $\hat{H}_{eff}$ and the kick operator $\hat{K}$.

\begin{equation}
\hat{H}_{eff}=\sum\limits_{n=0}^{\infty}\frac{1}{\omega^n}\hat{H}_{eff}^{(n)}, \;\; \hat{K}(t)=\sum\limits_{n=0}^{\infty}\frac{1}{\omega^n}\hat{K}^{(n)}(t)
\end{equation}

The Hamiltonian $\hat{H}$ in equation (\ref{eq:H - Supplement}) is of this form with

\begin{align}
   & &\hat{H}_0&=\frac{p^2}{2m}-\frac{1}{2}\delta_{RF}+\Omega\hat{\sigma}_x, &\\
   & &\hat{B}&=\frac{1}{2} k_0\hat{z}\hat{\sigma}_z, \;\;\qquad\qquad g(t)=\sin(\omega t+\phi_{RF}), &\\
   & &\hat{A}&=\Omega\sigma_x, \qquad\qquad\quad\quad f(t)=T\sum\limits_{n}\delta(t-nT)-1 &
\end{align}

Functions $f(t)$ and $g(t)$ meet the requirement of having zero mean value over a period $T$.

The kick operator is in 0-th order:
\begin{equation}
\hat{K}^{(0)}=\hat{B}G(t), \qquad G(t)=\omega\int\limits^{t}g(\tau)d\tau=-\cos(\omega t +\phi_{RF})
\end{equation}
The effective Hamiltonian to the lowest order in $1/\omega$ can be expanded as
\begin{equation}
\hat{H}_{eff}=\hat{H}_0+\sum\limits_{n=1}\frac{i^n}{n!}\overline{G^nf}\underbrace{[B...[B,}_\text{n}A]]+\sum\limits_{n=1}\frac{i^n}{n!}\overline{G^n}\underbrace{[B...[B,}_\text{n}H_0]]+O(1/\omega)
\end{equation}
After calculating all commutators and time-averaged coefficients before them, and grouping the terms proportional to $\hat{\sigma}_x$ and $\hat{\sigma}_y$, the expansion reduces to

\begin{equation}
\hat{H}_{eff}=\frac{\hat{p}^2_z}{2m}-\frac{1}{2}\delta_{RF}\hat{\sigma}_z+\Omega\cos(k_0z\cos\phi_{RF})\hat{\sigma}_x+\Omega\sin(k_0z\cos\phi_{RF})\hat{\sigma}_y+\frac{1}{16}\frac{k_0^2}{m}+O(1/\omega)
\end{equation}

The resulting effective Hamiltonian and micromotion operator are in exact agreement with the above equations (\ref{eq:H_eff - suppl}),
(\ref{eq:H_eff matrix - suppl}) and (\ref{eq:K- Suppl}).

If we apply an additional spatially-dependent unitary transformation
$\hat{R}_{z1}=\exp\left[-izk_{0}\cos\phi_{RF}\hat{\sigma}_{z}/2\right]$
corresponding to the choice $C=0$ of the intergration constant in
Eq. (\ref{eq:R_z - Supplement}), the transformed Hamiltonian becomes
translationally invariant and acquires the standard form of $H_{SOC}$
for one-dimensional spin-orbit coupling:
\begin{equation}
\hat{H}_{SOC}=\frac{1}{2m}(\hat{p}_z-\frac{1}{2}k_0\cos\phi_{RF}\hat{\sigma}_z)^2+\Omega\hat{\sigma}_x-\frac{\delta_{RF}}{2}\hat{\sigma}_z,
\end{equation}
where the spin-orbit coupling strength is described by the momentum shift $k_{0}\cos\phi_{RF}/2$. With the new kick operator
\begin{equation}
\hat{K}_{SOC}\left(t\right)=-k_{0}z\hat{\sigma}_{z}\left[\cos\left(\frac{2\pi}{T}t+\phi_{RF}\right)-\cos\phi_{RF}\right]\label{eq:K-SOC- Suppl-1}
\end{equation}
the time evolution can be written as
\begin{equation}
U\left(t_{2},t_{1}\right)=e^{-i\hat{K}_{SOC}\left(t_{2}\right)}e^{-i\hat{H}_{SOC}\left(t_{2}-t_{1}\right)}e^{i\hat{K}_{SOC}\left(t_{1}\right)}.\label{eq:evolution-SOC - Suppl}
\end{equation}

In that case the operator $\hat{K}_{SOC}\left(t\right)$ has a non-zero
temporal average, so it cannot be treated as a pure micromotion operator.
Similarly $\hat{H}_{SOC}$ can not be considered as an effective Hamiltonian
for the time-periodic Hamiltonian (\ref{eq:H - Supplement}). It is rather a Hamiltonian related
to the true effective Hamiltonian by the unitary transformation: $\hat{H}_{SOC}=\hat{R}_{z1}^{\dagger}\hat{H}_{eff}\hat{R}_{z1}$.
Note that ref. \cite{SLuo-supp} has obtained $\hat{H}_{SOC}$ as a stroboscopic Floquet Hamiltonian
after applying a unitary transformation of the form (\ref{eq:R_z - Supplement}) with $C=0$.

\end{document}